# Charge mechanism of low-frequency stimulated Raman scattering in virus suspensions

V. B. Oshurko[1,2], O. V. Karpova[3], M. A. Davydov[1], A.N.Fedorov[1], A. F. Bunkin[1], S. M. Pershin[1], M. Y. Grishin[1]

[1] *Prokhorov General Physics Institute of the Russian Academy of Sciences*
[2] *Moscow State Technological University Stankin*
[3] *Lomonosov Moscow State University*

**Abstract:** A simple physical mechanism of stimulated light scattering on viruses in a water suspension—similar to Langmuir wave mechanism in plasma—is proposed. The proposed mechanism is based on the dipole interaction between the light wave and the inevitable uncompensated electrical charge on a virus in a water environment. Experimental data on tobacco mosaic virus are presented to support the proposed physical mechanism. It is demonstrated that stimulated amplification spectral line frequencies observed experimentally are well explained by the proposed mechanism. In particular, the absence of lower-frequency lines and the shifting of generation lines when the pH changes are due to ion friction in the ionic solution environment. The selection rules observed experimentally also confirm the dipole interaction type. It is shown that microwave radiation on virus acoustic vibrations frequency should appear under such scattering conditions. We demonstrate that such conditions also allow for local selective heating of viruses from dozens to hundreds of degrees Celsius. This effect is controlled by the optical irradiation parameters and can be used to selectively affect a specific type of virus.

**Introduction**

Recently, many works have been carried out on direct observation of acoustic vibrational modes of nanoscale objects (in particular, viruses) by means of low-frequency optical Raman scattering [01] and extraordinary acoustic Raman scattering (EAR) [02,03]. However, these techniques allow to study only single particles placed in artificial environment, e.g. on a surface or using double nanoholes [03], thus limiting practical applications. A conventional electrostriction mechanism is suggested for all these cases, but there are many publications concerning low-frequency stimulated Raman scattering on viruses and nanoparticles in water environment [1-3] that are based on a theory of a stronger excitation mechanism analogous to Langmuir waves in plasma. As it turns out, this mechanism explains several ambiguous features of such light scattering. Obviously, of greatest interest are nanoscale biological objects like viruses in water suspensions [4]. There is no doubt that gaining information about such objects by a new technique that also provides an opportunity to affect these objects is important for biomedical applications. The suggested new mechanism opens a possibility of selectively affecting certain types of viruses by optical means, namely biharmonic pumping.

At the same time, the mechanism of stimulated light scattering is not completely clear for such objects. Stimulated light scattering with frequency shift corresponding to acoustic vibrations frequency characteristic for nanoscale objects is traditionally explained by models based on electrostriction (ponderomotive) mechanism. This mechanism suggests the modulation of electric susceptibility of an electrically neutral medium by an external field [3] which clearly does not fit the discussed case. It is known that biological and non-biological objects in water suspensions (e.g. viruses and polymeric nanospheres, respectively) have a significant non-compensated electrical charge [5] and virus' protein shell (capsid) composed of amino acids undergoes electrolytic dissociation in water environment [6]. This explains the phenomenon of proteins and ribonucleic acids electrophoresis caused by non-compensated electrical charge as the number of amino acid residues or nucleic bases is different from the number of alkaline residues [6]. Consequently, proteins and viruses demonstrate a so called isoelectric point, i.e. such pH of the environment when the non-compensated charge vanishes. It means that by varying the pH of the solution one can control the virus non-compensated charge value.

The published experimental works show that polymeric nanospheres in water suspensions also possess a significant charge, e.g. 20e (elementary charges) per single 50-nm polystyrene nanoparticle [5]. It is obvious that the energy of electrostatic interaction between such charge and the electric field in the laser beam waist is many orders of magnitude larger that the electrostriction interaction energy (i.e. the energy of interaction between the induced dipole moment and the external field). The electrostatic force affecting the charge is evidently much stronger than electrostriction forces. Until now, this scenario has not been taken into account, although the interaction between the light beam electromagnetic field and the charge may be the main mechanism determining low-frequency stimulated light scattering phenomenon.

Moreover, the existence of a significant non-compensated electrical charge leads to the strong suppression of acoustic vibrations in such particles surrounded by water due to so called ionic (or cataphoretic) friction [5]. In other words, the friction should be so strong that all vibrations were virtually impossible, however, the experiments show that exactly in water suspensions a directional stimulated emission is formed on Stokes frequency (i.e. red-shifted by the value of acoustic vibrations frequency inherent to these particles). Here and on we will call this effect the "low-frequency stimulated light scattering" (LFSS).

There are a series of LFSS peculiarities, which are comparable to conventional stimulated Raman scattering (SRS). For example, unlike SRS, LFSS is almost never observed on the frequency of the lowest vibrational mode [4-6]. Furthermore,—as will be shown later—in the case of viruses, the observed LFSS frequency can depend on the ion and virus concentrations in the solution.

LFSS also differs noticeably from the stimulated Brillouin scattering (SBS) process. In SBS, the stimulated light scattering appears on an acoustic wave inside a macroscopic object, and the frequency shift is defined only by the speed of sound in this object. In LFSS, the frequency shift is clearly caused by the scattering of acoustic waves on nanoscale objects and is dependent on characteristics of the nanoscale objects, such as size.

Thus, these peculiarities and the unclear mechanism of stimulated light scattering on nanoscale objects' acoustic vibrations permits stating that LFSS is a new nonlinear optics phenomenon.

This study develops a model that describes the LFSS phenomenon based on a charge (or Langmuir wave) mechanism and explains the experimentally observed features of the process.

**Charge mechanism of LFSS**

Consider a simplified model of a charged dielectric nanoscale object (a nanosphere or nano-sized virus particle) in the electromagnetic field of a light wave. The object itself is not electrically neutral in this case. Usually, the excess charge of such an object is compensated for by oppositely charged ions in the solution or a surfactant, which is always added to a nanoparticle suspension to prevent adhesion. This charge-compensating ionic shell size is defined by the Debye length for the solution. The existence of such a shell requires adding another factor to the model, i.e., the friction that inevitably appears during acoustic vibrations of a nanoscale object in an ionic shell.

Let the plane-polarized light wave be directed along the z axis and the electric field vector oscillate along the x axis. The Maxwell equation for light wave $E(r,t)$ is given as

$$\widehat{D}E(r,t) = \nabla(\nabla E(r,t)) + \frac{4\pi}{c^2}\frac{\partial J(r,t)}{\partial t} \qquad (1)$$

where $\widehat{D} = \Delta - 1/c^2 \partial^2/\partial t^2$ is the d'Alembertian, $c$ is the speed of light, and $J(r,t)$ is the density of the current induced by the acoustic movement of the charge. In the case of regular SRS or the electrostrictional (ponderomotive) mechanism, the first term on the right side of the equation equals zero because of the zero total charge of the medium. However, this is not the case being considered, and the field divergence equals the charge density

$$\nabla E(r,t) = 4\pi\rho(r,t)$$

where $\rho(r, t)$ is the charge volume density.

Let the object be charged uniformly with the initial charge density constant over the volume ($q_0 = Const$). If the object is absolutely rigid (its elastic modulus equals infinity), there will be no movement apart from reciprocating oscillations in the field $\vec{E}$ of the light wave. Such oscillations make no contribution to inelastic light scattering and will not be considered. Acoustic vibrations are possible in an elastic object with a finite Young's modulus and can appear spontaneously at room temperature because the acoustic phonon energy $\hbar\Omega$ is much less than the heat energy $k_B T$.

Let us consider the acoustic displacement $u = r' - r$, where $r'$ is the displaced position of point $r$ in an acoustic wave. The solid body acoustic equation can be expressed as

$$\Delta \boldsymbol{u}(r,t) - \Gamma \frac{\partial \boldsymbol{u}(r,t)}{\partial t} - \frac{1}{v^2}\frac{\partial^2 \boldsymbol{u}(r,t)}{\partial t^2} = \frac{1}{v^2}\boldsymbol{F}_m \qquad (2)$$

where $v$ is the speed of sound, and $\boldsymbol{F}_m$ is the force mass density, i.e., volume density ($\boldsymbol{F}_v = d\boldsymbol{F}/dV$) divided by the density of the mass, $\boldsymbol{F}_m = \boldsymbol{F}_v/\rho$. Here, we also introduce the phenomenological term that describes the friction proportional to velocity with the coefficient $\Gamma$. As it was mentioned earlier, this is necessary to account for external ionic shell resistance to oscillations.

It is obvious that in our case, the force density is

$$\boldsymbol{F}_m = q(r,t)/\rho(r,t)\boldsymbol{E}(r,t)$$

where $q$ is the charge volume density.

An acoustic wave in a nanoscale object can be naturally considered as a wave of density modulation $\rho'(r,t) = \rho_0 f(r,t)$, where $f$ is the dimensionless function describing this wave. The charge density modulation is described by the same function $f$ and consequently, the ratio $q'/\rho'$ in the right side of Eq. (2) remains constant.

Then the only form of direct interaction between the light wave field and the nanoscale object charge is *charge decompensation* caused by the well-known Debye–Falkenhagen effect [11]. When an electromagnetic wave passes through an ionic solution, starting from a wave frequency, the ion *heavy* hydration shell fails to *keep up* with the ion movement, leading to a sharp growth in solution conductivity. For the majority of water solutions, this frequency lies in the region of several gigahertz and is lower than the observed characteristic frequencies of nanosphere/virus acoustic vibrations (dozens of gigahertz). In recent research, the inertia of hydration shells is always taken into account [12] in explaining the emergence of uncompensated charge while exciting acoustic vibrations in charged nanoscale objects.

The *charge decompensation* wave is analogous to the well-known Langmuir waves in plasma [13], with the only difference being that the characteristics of the *charge decompensation* wave depend on the mechanical properties of the nanoscale object.

For simplicity, consider a one-dimensional problem describing a tobacco mosaic virus particle that looks like a thin cylinder. Like the light wave $E$ vector, let the cylinder axis be directed along the $x$ axis. If the negative (or positive) charge volume density of the virus is modulated by an acoustic wave $q^-(x,t) = q_0^- f(x,t)$, then the opposite ion charge density in the hydration shell in the solution remains constant ($q^+ = q_0^+$) due to the Falkenhagen effect and equals $q_0$ for the one-dimensional problem. The total charge density is then defined by the difference $q(x,t) = q_0^- f(x,t) - q_0^+$. An easy direct calculation shows that the total charge density is $q(x,t) = q_0 \frac{\partial u}{\partial x}$.

Apart from the force induced by the light wave field $q_0 u_x E(z,t)$, there are also interactive

forces between charged areas similar to that in Langmuir waves in plasma. These forces are

$$F_e = q(x,t)E' = q(x,t)\int \frac{\partial E'}{\partial x}dx = q(x,t)\int 4\pi q(x,t)dx = 4\pi q_0^2 u u_x \qquad (3)$$

where $E'$ is the field created by uncompensated charges.

Finally, the equation for acoustic displacement $u(x,t)$ is

$$u_{xx} - \Gamma u_t - \frac{1}{v^2}u_{tt} = \frac{q_0}{\rho_0 v^2}u_x E(z,t) + 4\pi \frac{q_0^2}{\rho_0 v^2}u u_x \qquad (4)$$

**Dipole approximation**

It is difficult to find a general solution of this equation for a harmonic light wave. However, it is evident that a charge moving in the field of a light wave does not itself cause inelastic scattering. To find the inelastic components of the scattering, it is necessary to examine the processes in which the light wave field modulates a parameter of the system. In particular, let us consider the modulation of the total dipole moment by the external electric field. To achieve this, an auxiliary problem has to be solved: consider a constant force affecting the dipole moment of the object instead of a light wave field, i.e., let us determine how a constant external field changes the dipole moment.

In the right side of Eq. (4), there is the density of the force affecting the object via the field and the interaction between the areas of the distributed charge. The estimation readily shows that although the LFSS field is relatively weak compared to the left side of Eq. (4), the second term in the right side of Eq. (4) (the interaction between the areas) is much smaller than the first term for realistic acoustic vibration amplitudes and can thus be neglected.

It is obvious that in the absence of acoustic vibrations, the charge distribution is uniform and there is no dipole moment—or, to be precise, it is possible to choose a reference frame in which the dipole moment equals zero. The dependence of a charged object dipole moment on the chosen reference frame is insignificant as force is defined only by the dipole moment derivative. When an acoustic wave appears, charged areas are created due to charge decompensation, and the total dipole moment (per unit volume) for a virus of length $L$ and its derivative are

$$p = \frac{q_0}{L}\int_0^L x u_x dx \qquad (5)$$

and

$$p_x = \frac{q_0}{L}L u_x(L) = q_0 u_x(L)$$

In dipole approximation, to determine the force affecting each element $dx$ at point $x$, we determine the coordinate-dependent dipole moment $p(x)$ and its derivative as

$$p_x(x) = q_0 u_x$$

The light wave electric field force density is

$$F_v = \frac{1}{2}p_x E(z,t) = \frac{1}{2}q_0 u_x E(z,t) \qquad (6)$$

Estimations show that the magnitude of the dipole moment and the force affecting the dipole are in fact much larger than the electrostriction (ponderomotive) forces traditionally considered in nonlinear optics.

In Eq. (5) and Eq. (6), the displacement $u(x,t)$, and consequently, the dipole moment, are a function of the external field E. To consider the dependence $p(E)$ using the simplified model, let us examine a charged object (a virus) in a constant external field $E$. As $E = Const$ in the right side of Eq. (4), using the standard substitution $u(x,t) = U(t)V(x)$, we derive the following system of equations

$$U_{tt}(t) + \gamma U_t(t) + \Omega_m^2 U(t) = 0 \qquad (7)$$
$$V_{xx}(x) - \frac{q_0}{2\rho_0 v^2}EV_x(x) + \frac{\Omega_m^2}{v^2}V(x) = 0 \qquad (8)$$

Here, we denote $\gamma = \Gamma v^2$ and the common constant $-\Omega_m^2/v^2$. The solution of Eq. (8), with the free ends condition $u_x(0) = u_x(L) = 0$, is

$$V(x) = u_0 e^{\frac{q_0 E x}{4\rho_0 v^2}} \cos\left(x\sqrt{\frac{\Omega_m^2}{v^2} - \frac{E^2 q_0^2}{16\rho_0^2 v^4}}\right) \qquad (9)$$

where $\Omega_m = m\pi v/L$ is the $m^{\text{th}}$ mode eigenfrequency, and $u_0$ is the amplitude. Hence, we can determine the derivative of the dipole moment based on Eq. (5), Eq. (6), and Eq. (9)

$$p_x(E) = U(t)\frac{u_0 q_0}{4\rho_0 v^2} e^{\frac{q_0 E L}{4\rho_0 v^2}}(q_0 E \cos(\Omega_E L) - (4v^2\rho_0\Omega_E)\sin(\Omega_E L)) \qquad (10)$$

where

$$\Omega_E = \sqrt{\frac{\Omega_m^2}{v^2} - \frac{E^2 q_0^2}{16\rho_0^2 v^4}} \qquad (11)$$

In Eq. (10) and Eq. (11), $U(t)$ is the solution of Eq. (7) in the form of a periodic function with $\sqrt{\Omega_m^2 - \gamma^2}$ frequency and the attenuation coefficient $\gamma$. For our purposes, let $U = 1$. In Eq. (7) and Eq. (8), the field $E$ is constant, and Eq. (7), Eq. (8), and Eq. (9) are not intended to solve the stimulated amplification problem but are only used to estimate the dipole moment dependence on the field.

Under the conditions of LFSS, the field can be considered small, i.e., $E \ll 4\rho_0\Omega_m v/q_0$. This is true up to $E: 10^9$ V/m, which is significantly stronger than the intensities used in experiments, i.e., $I: 1$ MW/cm². Thus, $p_x(E)$ can be expanded into a series using $E$

$$p_x(E) = \frac{q_0^2 u_0 (-1)^m}{4\rho_0 v^2} E + \frac{3}{32}\frac{q_0^3 u_0 L(-1)^m}{\rho_0^2 v^4} E^2 + \ldots \qquad (12)$$

The series coefficients depend on the amplitude of the initial oscillation $u_0$, which is quite different in regular molecular SRS. In fact, almost any molecule has non-zero polarizability; however, in our case, the polarizability is not equal to zero only if the initial charge distribution is non-uniform. This is possible, for example, due to spontaneous acoustic vibrations. If there is no non-uniform distribution, then the polarizability (i.e., the dipole moment dependence on the external field), evidently equals zero. The external field itself affects all areas equally, and hence, does not create a non-uniform charge distribution, which would result in a non-zero dipole moment.

**Stimulated amplification of radiation**

Given the dipole approximation, we substitute the force density with $1/2 p_x(E(z,t))E(z,t)$ on the right side of the acoustic equation, Eq. (4). Accounting for the expansion of Eq. (12), for the first two terms of the series we obtain

$$u_{xx} - \Gamma u_t - \frac{1}{v^2} u_{tt} = \alpha_1 E(z,t)^2 + \alpha_2 E(z,t)^3 \qquad (13)$$

where

$$\alpha_1 = \frac{q_0^2 u_0 (-1)^m}{2\rho_0 v^2}, \alpha_2 = \frac{3q_0^3 u_0 L(-1)^m}{16\rho_0^2 v^4} \qquad (14)$$

and $E(z,t)$ is the light field. To determine the amplification of the radiation components in a stimulated process, we consider a combination of a pump wave and a Stokes wave based on the traditional nonlinear optics approach $E(z,t) = E_0 \exp(i\omega_0 t - ik_0 z) + E_s \exp(i\omega_s t - ik_s z) + c.c.$ In this case, the right side of Eq. (13) does not depend on $x$, and Eq. (13) can be split into two ordinary equations

$$U_{tt}(t) + \gamma U_t(t) + \Omega_m^2 U(t) = \alpha_1 E(z,t)^2 + \alpha_2 E(z,t)^3 \qquad (15)$$

and

$$V_{xx}(x) + \frac{\Omega_m^2}{v^2} V(x) = 0 \qquad (16)$$

The second equation, Eq. (16), with the free ends condition is an expression for eigenfrequencies $\Omega_m = m\pi v/L, m = 1,2,..$

$$V(x) = u_0 \cos\left(\frac{\Omega_m}{v} x\right) \qquad (17)$$

In the first equation, Eq. (15), we substitute the combination of pump and Stokes waves (frequencies $\omega_0$ and $\omega_s$, respectively) mentioned earlier and derive a solution $u = UV$ at a frequency $\Omega = \omega_0 - \omega_s$

$$u_\Omega = \alpha_1 V(x) \frac{E_0 E_s}{\Omega_m^2 - \Omega^2 + 2i} e^{(i\Omega t - i\,\Omega z)} \tag{18}$$

Similarly, we derive a solution at the Stokes frequency $\omega_s$

$$u_s = \alpha_2 V(x) \frac{E_0^2 E_s}{\Omega_m^2 - \omega_s^2 + 2i\gamma\omega_s} e^{(i\omega_s t - i k_s z)} \tag{19}$$

It can be seen that the second solution for the Stokes component $\omega_s$ does not have pronounced resonances at detuning equal to eigenfrequencies $\omega_0 - \omega_s = \Omega_m$ and does not explain the observed stimulated amplification at these frequencies.

Consequently, the only possible stimulated amplification mechanism for Stokes or anti-Stokes frequencies is a simultaneous amplification at a microwave frequency $\Omega$ (at which waves with frequencies $\omega_0$ and $\omega_s$ are combined) and further amplification of the frequency $\omega_s$ when the frequencies $\omega_0$ and $\Omega$ are combined.

**Coupled waves equations**

To investigate the possibility of such simultaneous amplification, we consider the propagation of four waves at the frequencies $\omega_0, \Omega, \omega_s,$ and $\omega_a$ (where $\omega_a$ is the anti-Stokes radiation frequency). Again, we derive the solutions of Eq. (13) for these frequencies considering only the first order terms on the right side of the equation

$$u_0 = \alpha_1 V(x) \frac{(E_0 E_s + E_0 E_a)}{\omega_m^2 - \omega_0^2 + 2i\gamma\omega_0} e^{(i\omega_0 t - i k_0 z)} \tag{20}$$

$$u_s = \alpha_1 V(x) \frac{E_0 E_\Omega}{\Omega_m^2 - \omega_s^2 + 2i\gamma\omega_s} e^{(i\omega_s t - i k_s z)} \tag{21}$$

$$u_a = \alpha_1 V(x) \frac{E_0 E_\Omega}{\Omega_m^2 - \omega_a^2 + 2i\gamma\omega_a} e^{(i\omega_a t - i k_s z)} \tag{22}$$

$$u_\Omega = \alpha_1 V(x) \frac{(E_0 E_s + E_0 E_a)}{\Omega_m^2 - \Omega^2 + 2i\gamma\Omega} e^{(i\Omega t - i\,\Omega z)} \tag{23}$$

Next, to substitute these equations into the wave equation, Eq. (1), we define the terms $J_t = \mu_0 \partial^2 u_i / \partial t^2$, and using slowly varying envelope approximation, we obtain a system of bound waves equations in which we wave out dependence on the $x$ coordinate

$$\frac{dE_0}{dz} = \beta_0 E_\Omega (E_s + E_a), \frac{dE_s}{dz} = \beta_s E_\Omega E_0$$
$$\frac{dE_a}{dz} = \beta_a E_\Omega E_0, \frac{dE_\Omega}{dz} = \beta_\Omega E_0 (E_s + E_a) \tag{24–27}$$

where $\beta$ are constants defined by equations (20)–(23). Solving these equations for the Stokes component, we obtain

$$E_s(z) = \frac{E_{s0}}{1 - E_{s0} Re(\chi) z} \tag{28}$$

where $E_{s0}$ is the magnitude of Stokes radiation (e.g., spontaneous Stokes radiation) at $z = 0$, and $\chi$ is

$$\chi(\Omega) = \sqrt{-\frac{c^2\mu_0{}^2\alpha_1{}^2\omega_0\Omega}{(\Omega_m{}^2-\omega_0{}^2+2i\gamma\omega_0)(\Omega_m{}^2-\Omega^2+2i\gamma\Omega)}}$$

$$\left(\frac{(\omega_0+\Omega)\left(\Omega_m{}^2-(\omega_0-\Omega)^2+2i\gamma(\omega_0-\Omega)\right)}{\left(\Omega_m{}^2-(\omega_0+\Omega)^2+2i\gamma(\omega_0+\Omega)\right)(\omega_0-\Omega)}+1\right)$$
(29, 30)

**Experimental results**

We can now compare the developed theoretical model with experimental results on low-frequency stimulated light scattering in suspensions of tobacco mosaic virus (TMV) in tris buffer and polystyrene nanoparticles of various diameters (70 nm, 100 nm, and 500 nm) and concentrations in distilled water.

In the experimental setup described in detail in the Bunkin et al. study [7], the cuvette with the studied suspension was irradiated by focused second harmonic pulses of a single-frequency YAG:Nd$^{3+}$ laser (wavelength $\lambda$ of 532 nm, pulse duration $\tau$ of 10 ns, and pulse energy $E_p$ of up to 40 mJ). The radiation was focused onto the cuvette center by a lens with a focal length f of 30 mm. Stimulated light scattering spectra were studied using a Fabry-Pérot interferometer and recorded with a complementary metal-oxide semiconductor (CMOS) camera, a Basler acA1920-40um.

The results of the TMV suspension spectra measurements [15] are presented in Fig. 1. For a virus concentration of $1.0 \cdot 10^{12} cm^{-3}$ and a laser pulse energy $E_p$ of ~20 mJ, an LFSS line was detected at a 1.47 cm$^{-1}$ shift relative to the pump line, which corresponds to a 44.1 GHz oscillation frequency (Fig. 1a). For a virus concentration of $2.0 \cdot 10^{12} cm^{-3}$ and a laser pulse energy of ~30 mJ, an LFSS line was detected at a 1.046 cm$^{-1}$ shift (31.38 GHz, Fig. 1b). The experimental results also confirm this directional character of the LFSS emission. An electron microscopy view of a dried TMV suspension after LFSS experiments is shown in Fig. 2.

**Vibrational states selection rules**

Considering the experimentally measured values of the Young's modulus and speed of sound of TMV in the study by Stephanidis et al. [14], it is easy to calculate acoustic vibrations frequencies for a TMV cylinder 300 nm in length and 18 nm in diameter in a water environment. Numerical calculations performed in the COMSOL Multiphysics software for all possible vibration types (supposing speed of sound $v = 3430$ m/s, density $\rho_0 = 1100$ kg/m³, and Young's modulus E = 9.5 MPa [7]) yield the results in Table 1.

Table 1. TMV virus longitudinal vibration mode frequencies.

| m | 0 | 1 | 2 | 3 | 4 | 5 | 6 | 7 |
|---|---|---|---|---|---|---|---|---|
| f (GHz) | 5.71 | 11.43 | 17.15 | 22.86 | 28.58 | 34.30 | 40.01 | 45.73 |

| m | 8 | 9 | 10 | 11 | 12 | 13 | 14 | 15 |
|---|---|---|---|---|---|---|---|---|
| f (GHz) | 51.45 | 57.16 | 62.88 | 68.60 | 74.32 | 80.04 | 85.76 | 91.48 |

Bending vibration frequencies turned out to be higher than 112 GHz, and radial vibration frequencies are above 190 GHz.

As can be seen, the longitudinal vibration frequencies 34 GHz and 45 GHz coincide well with the observed TMV LFSS frequencies (31 GHz and 44 GHz). Notably, increasing the virus concentration in tris buffer in the experiments yielded the observed LFSS frequency switch from 44 GHz to 31 GHz. In one experimental session, the 44 GHz and 31 GHz lines even appeared simultaneously, but the intermediate frequency, ≈40 GHz, was never observed.

This observation is in good agreement with the proposed theoretical model. In fact, the ≈40 GHz frequency corresponds to an even vibrational mode for which the integral (Eq. (5)) defining the dipole moment (Eq. (17)) equals zero. Hence, the experiment confirms the dipole type of interaction and the subsequent selection rule for odd vibrational modes.

Therefore, the absence of lower vibrational frequencies in the amplified LFSS lines can now be explained. In fact, the expression in Eq. (29) denoting the gain increment defines only one *window* of amplification. As can be seen from Eq. (29), this window is dependent on the magnitude of the ionic friction $\gamma$. Fig. 3 presents the gain increment $Re(\chi)$ as a function of $\Omega$ for a set of resonance odd mode frequencies $\Omega_m$ from Table 1 ($m = 1,3,..$) for two values of the ionic friction coefficient (2.0 GHz and 10.0 GHz). It is evident that the weaker friction (2.0 GHz) results in a maximal gain increment $Re(\chi)$ for $m = 4$ at a frequency of approximately 33 GHz. The apparent reason for this is non-zero ionic friction. The increase in ionic friction coefficient to 10.0 GHz—as expected—leads to an enormous broadening and merging of resonance peaks, with the maximum at ~41 GHz.

Fig. 5 and Fig. 6 illustrate the growth of these components with an increase in the interaction length $z$, based on Eq. (31). On reaching $z = 2.6$ mm, the spectrum is reduced to virtually a single line: the 41 GHz line for the 10.0 GHz friction coefficient, and the 33 GHz line for the 2.0 GHz friction coefficient. Fig. 4 shows the radiation spectra for $z = 2.6$ mm.

These results correspond well to the experimental data and allow for explaining the observed LFSS frequency change caused by an increase in the TMV concentration in the suspension. The ionic friction evidently decreases with a decrease in the number of ions *compensating* the charge on the virus. During the experiments, we increased the virus concentration by evaporating water from the solution, but *heavy* virus particles and tris molecules, which create a weak-alkaline environment, were not evaporated. Hence, the pH of the solution increased. As mentioned earlier, the isoelectric point (i.e., the pH for which the virus becomes electrically neutral) lies at rather high pH values (i.e., high alkalinity). Therefore, as water evaporates, the pH increases and the value of the uncompensated charge decreases. Consequently, there are less solution ions compensating for the charge of the virus, leading to a reduction in ionic friction, which in turn results in a change in amplification frequency (Fig. 4–Fig. 6).

Only one adjustable parameter was used to describe this LFSS phenomenon, i.e., the ionic friction magnitude. Notably, this phenomenon cannot be explained in terms of other physical mechanisms.

**Nanoscale objects heating**

Another conclusion follows directly from the preceding conclusion. If some part of the pump energy is spent on overcoming friction, this should result in the heating of the local nanoscale particles. To roughly estimate this heating in terms of the described theoretical model, we suppose that the virus is a dipole with a moment $p \approx QL$, where $Q$ is the uncompensated charge value experimentally estimated as $\sim 50 \cdot 1.6 \cdot 10^{19}$ C. The energy of the dipole in the field $E$ is then $A \approx QLE$, and the total power can be taken as $P \approx QLE\Omega_m$. The portion of this power converted into heat can be estimated from the magnitude of the ionic friction as $(1 - exp(-\gamma\tau_p)$, where $\tau_p$ is the laser pulse duration. With the friction coefficient $\gamma \approx 10^9 c^{-1}$ and pulse duration $\tau_p \gg 1/\gamma$, all of the absorbed radiation is transformed into heat. The amount of heat produced during the pulse $\tau_p$ is $A \approx QLE\Omega_m\tau_p$. Then, if we neglect the thermal conductivity (during the laser pulse, the heat wave diffusion length is ~100 nm, given a water thermal conductivity $\chi = 1.9 10^{-5} cm^2/s$), we obtain the final heating temperature as

$$T \approx \frac{QL\Omega_m\tau_p}{\rho_0 C_p V} E_0 \qquad (31)$$

where $C_p \approx 3000 J/(kgK)$ is the protein heat capacity. Then, at a light field strength $E_0 \approx 10^8 V/m$, we obtain the heating as $T \approx 26.0 K$ during a 10 ns laser pulse. This rough estimate shows that heating can occur in experiments, though the temperature is still insufficient to denature the proteins.

However, it is obvious that a simple several-fold pulse duration increase (up to 0.1–1 μs) may lead to a strong local heating of up to dozens or hundreds of degrees Celsius. Nevertheless, the overall heating may be insignificant in all these cases. Notably, the virus is heated by local selective non-resonant radiation that is not absorbed by the suspension. This is the stimulated nature of the process that provides highly selective heating of certain virus types.

It is known that TMV heating up to only 94 °C leads to partial denaturation of capsid proteins and the formation of a spherical virus [17,18]. Therefore, selective heating of given virus types that changes the capsid shape may strongly affect the virus functioning.

It is significant that LFSS allows for selectively heating given types of nanoscale objects up to substantial temperatures. The very possibility of such heating opens new prospects for local selective treating of nanoscale objects, viruses, and cellular organelles via nonlinear optics for biomedical applications.

**Conclusion**

In this study, a simple physical mechanism of low-frequency stimulated light scattering on viruses in water suspensions is proposed. The proposed model differs significantly from the traditional stimulated scattering mechanism based on electrostriction. The model is based on a dipole interaction between the light wave and the uncompensated electrical charge that inevitably exists on a nanometer scale in the virus in a water environment. The experiments were conducted by observing low-frequency stimulated light scattering in tobacco mosaic virus water suspensions, and the data obtained supports the proposed model. The selection rules observed experimentally confirm the dipole type of the interaction. It has been demonstrated that stimulated amplification spectral line frequencies observed experimentally cannot be explained by traditional mechanisms but are well explained by the proposed charge mechanism. In particular, the absence of lower-frequency lines and the *window of amplification* are due to ion friction in the ionic solution environment. It has been shown that under these conditions, microwave radiation should appear at virus acoustic vibration eigenfrequency. We demonstrate that such conditions also allow for local selective heating of virus to dozens and hundreds of degrees Celsius. This effect is controlled by optical radiation parameters and can be used to selectively destroy a specific virus type among other viruses in a solution.

This work was partially supported by a grant of the Ministry of Science and Higher Education of the Russian Federation # 075-15-2020-912 and grant # FSFS-2020-0025.

**Figures**

Fig. 1(a). Low-frequency stimulated scattering (LFSS) spectrum for the tobacco mosaic virus concentration ~$1.0·10^{12}$ cm$^{-3}$, $\Delta v$ ~1.47 cm$^{-1}$ (~44.1 GHz). (b) LFSS spectrum for the tobacco mosaic virus concentration ~$2.0·10^{12}$ cm$^{-3}$, $\Delta v$~1.046 cm$^{-1}$ (~31.38 GHz).

Fig. 2. Electron microscopy picture of dried TMV suspension after LFSS experiments.

Fig. 3. Gain *increment* for the frequencies of 10 odd modes (m =1,3,5,...) from Table 1 for two friction coefficients: 2.0 GHz (red) and 10.0 GHz (green).

Fig. 4. Spectral lines at the exit from the interaction region z = 2.5 mm, with the frequencies of 10 odd modes (m =1,3,5,...) from Table 1 taken into account, for two friction coefficients: 2.0 GHz (red) and 10.0 GHz (green).

Fig. 5. Gain dependence on the interaction length z for the frequencies of 10 odd modes (m = 1,3,5,...) from Table 1 for the friction coefficient 2.0 GHz.

Fig. 6. Gain dependence on the interaction length z for the frequencies of 10 odd modes (m = 1,3,5,...) from Table 1 for the friction coefficient 10.0 GHz.

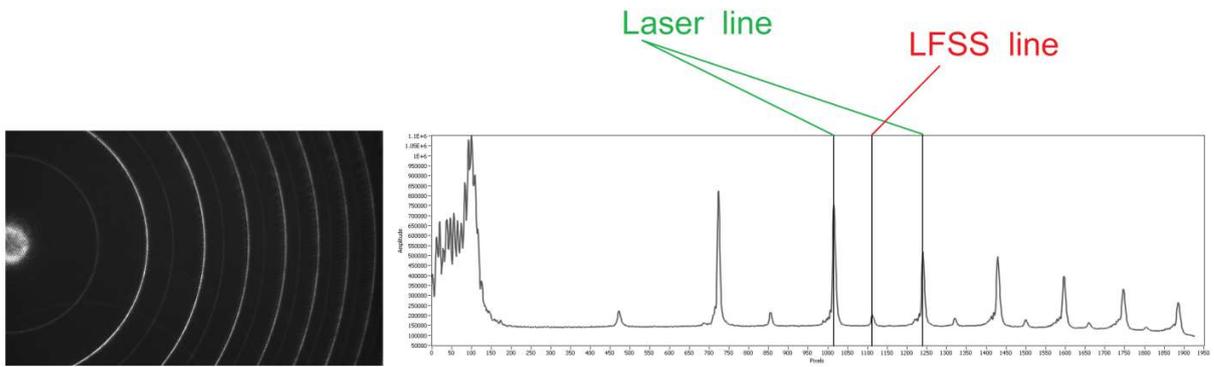

Fig. 1(a)

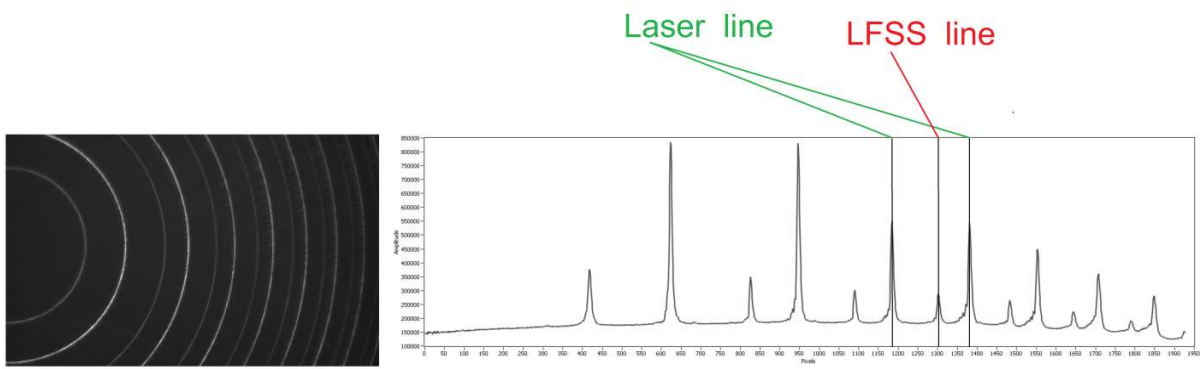

Fig. 1(b)

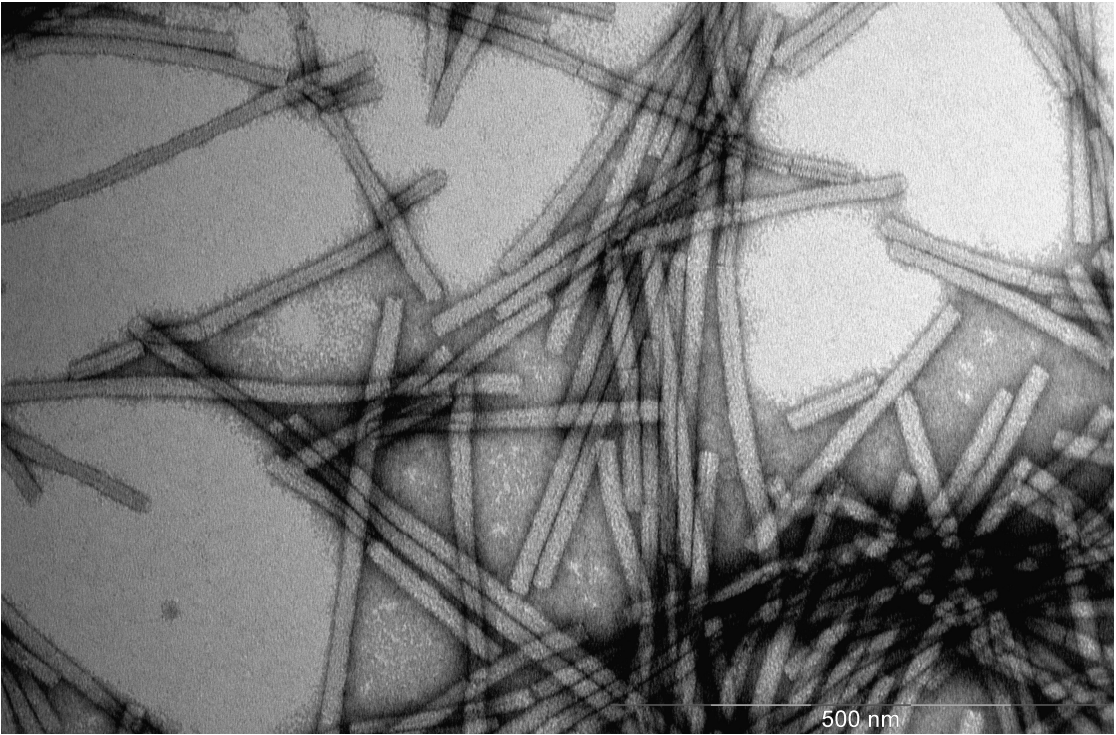

Fig. 2

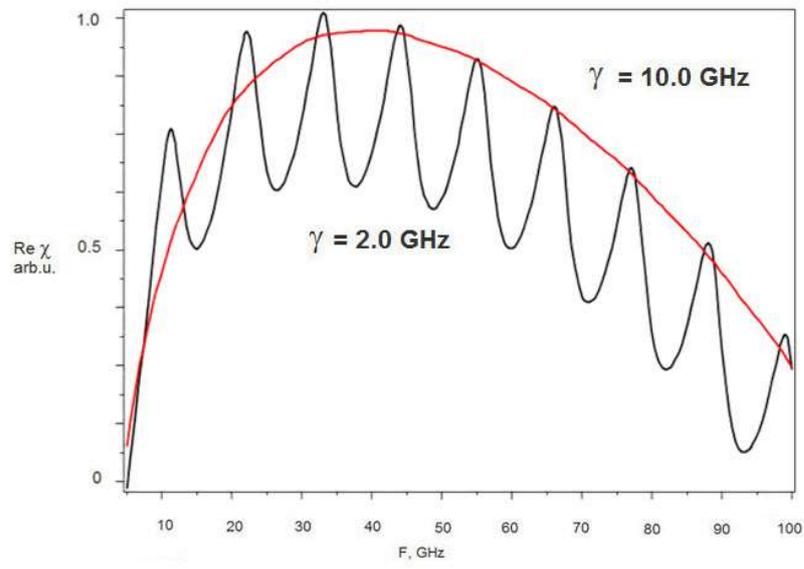

Fig. 3

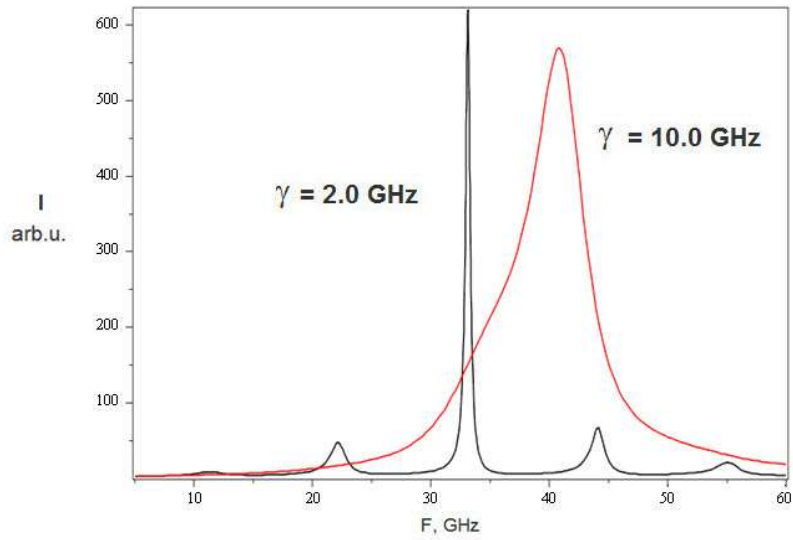

Fig. 4

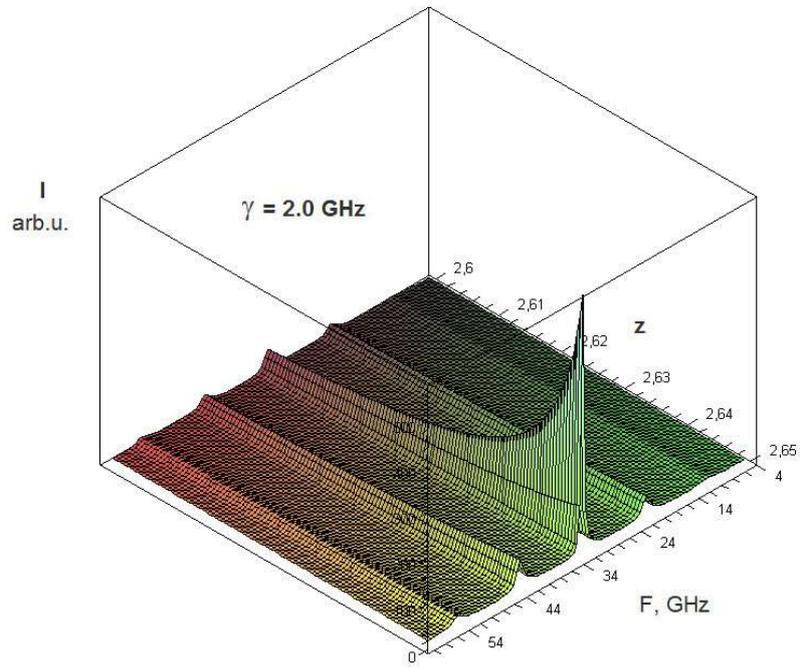

Fig. 5

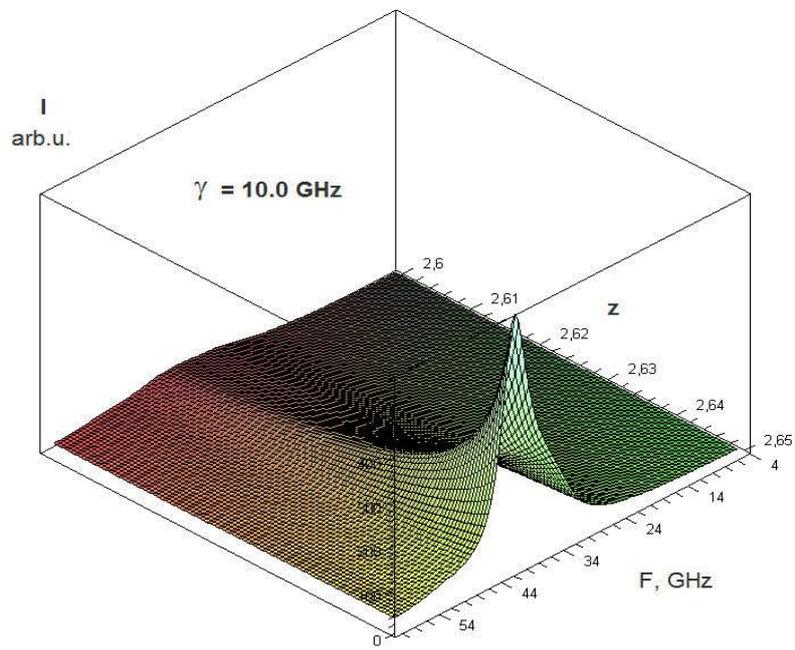

Fig. 6